\documentclass[pra,twocolumn]{revtex4}
\usepackage[T1]{fontenc}
\usepackage{amssymb,amsmath}
\usepackage{newtxtext}
\usepackage{color,graphicx,ulem}

\begin{document}

\title{Single-electron quantum dynamics in high-harmonic generation
spectrum from LiH molecule: 
analysis of potential energy surfaces for electrons
constructed from a model of localized Gaussian wave packets with
valence-bond spin-coupling}

\author{Koji Ando\footnote{E-mail: ando\_k@lab.twcu.ac.jp}}

\affiliation{Department of Information and Sciences, 
    Tokyo Woman's Christian University, 
    2-6-1 Zenpukuji, Suginami-ku, Tokyo 167-8585, Japan}

\begin{abstract}
High-harmonic generation (HHG) spectrum from a LiH molecule induced
by an intense laser pulse is computed and analyzed with potential
energy surfaces for electron motion (ePES) constructed from a model of
localized electron wave packets (EWP) with valence-bond spin-coupling.
The molecule has two valence ePES with binding energies
of 0.39 hartree and 1.1 hartree.
The HHG spectrum from 
an electron dynamics on the weaker bound valence ePES,
virtually assigned to
Li 2s,
exhibits a dominant peak
at the first harmonic without plateau and cut-off.
This compares with the free electron spectrum under oscillating laser field
and is comprehensive with the shape and depth of the ePES.
The other valence ePES, assingned to H 1s,
is deeper bound 
such that the overall
profile of the wave function is well approximated by a Gaussian
of the width comparable to the Li-H bond length.
However, a small fraction,
less than $10^{-3}$, of the probability
density amplitude tunnels out from the bound potential with high wave
number, and spreads over tens of nm's with parts recombining to the
molecule due to the laser field oscillation. 
This minor portion of the electronic wave function is the
major origin of the HHG extending up to 50
harmonic orders. 
Nonlinear dynamics
within the potential well induced by the laser field oscillation
also contributes to the HHG up to 30 harmonic orders.
\end{abstract}

\maketitle

\section{Introduction}

Study of electron dynamics in materials is an active field of
research 
reinforced
by recent advances of 
attosecond time-resolved
laser spectroscopy 
\cite{Bucksbaum2007,Kling2008,Krausz2009,Ramasesha2016}.
Electron dynamics induced by intense
and short laser pulse generates high harmonics of the input light
\cite{Krause1992,Schafer1993,Corkum1993,Baggesen2011,Bandrauk2013,
Itatani2004,Haessler2011,Salieres2012,Offenbacher2015,Ghimire2010,Yoshikawa2017}.
Proper analysis of the high-harmonic generation (HHG) spectrum is
expected to provide keys to understanding time-dependent behavior
of electrons in materials.
Current standard picture is the three-step model \cite{Corkum1993}
where 
half-ionized electrons 
driven by oscillating laser field 
recombine to the molecule
and emit light. 
This process is expected to self-probe 
the molecular electronic states via the so-called molecular
orbital tomography \cite{Itatani2004,Haessler2011,Salieres2012,Offenbacher2015}.
The three-step model has been generalized
to the momentum space for solid states \cite{Vampa2014}.

Computational modeling of electron dynamics in atoms and molecules 
cannot be a trivial extension of conventional 
molecular orbital or density functional methods. 
Ordinary atomic orbital basis functions are
insufficient to describe spatially large amplitude motion of
electrons induced by intense laser fields. Use of plane wave basis
functions will require many functions up to high wave numbers
to describe both local atomic and non-local ionized states.
Treating
many electrons will add further complexities
where adequacy of the mean-field orbital picture is not obvious.
Consequently, computational studies of 
electron dynamics under strong field on realistic
molecular models 
are rather scarce
(although the reference list herein
\cite{Grossmann2013,Geppert2008,Takemoto2011,Lostedt2012,
Telnov2013,Tolstikhin2013,Li2016,
Sato2013,Sato2015,Remacle2007,Nest2008,Remacle2011,Nikodem2017,Balzer2010,Ulusoy2012}
is not exhaustive). 
The size of the molecules that have
been treated so far is also limited.

In previous publications \cite{Ando2017,Ando2018}, 
we demonstrated that a model of floating
and breathing localized electron wave packets (EWP) with non-orthogonal
valence-bond (VB) spin-coupling 
\cite{Ando2009,Ando2012,Kim2014,Kim2014prb,Kim2015,Kim2016,Ando2016}
can be 
used to 
construct potential energy surfaces for single electron motion (ePES)
as functions of the EWP positions.
In Ref. \cite{Ando2018},  
the method was applied to a LiH molecule under short and intense laser pulse.
It was found that
a sum of the HHG
spectra from single electron dynamics on two valence ePES,
that were virtually assigned to 
Li 2s and H 1s,
agree well with that from the time-dependent
complete-active-space (TD-CASSCF) calculation \cite{Sato2013,Sato2015}. 
In this work, we continue the analysis
with further details of the wave functions
to clarify the origin and mechanism of the HHG spectrum.

Section \ref{sec:tc} outlines
the theory and computation.
The analysis of electron wave function
dynamics is presented in Sec. \ref{sec:results}. 
Section \ref{sec:concl} concludes.

\section{Model and Computation}
\label{sec:tc}

The theory and computation for 
the ePES with the VB EWP model
and for ther HHG spectra induced by an
intense laser pulse are mostly identical to those described in our previous
publication \cite{Ando2018}. 
We thus outline here the essential part,
with further details summarized in Appendix. 

In the VB EWP model, 
the total electronic wave function is represented by
a form of an antisymmetrized product of
spatial and spin functions. 
The spatial part is modeled by a product
of one-electron functions of a spherical Gaussian form
with variable central position \(\boldsymbol{q}_i\) and 
width \(\rho_i\),
\begin{equation}
    \phi_i(\boldsymbol{r})=(2\pi\rho_i^2)^{-3/4}\exp(-|\boldsymbol{r}-\boldsymbol{q}_i|^2/4\rho_i^2).
\label{eq:GWP}
\end{equation}
The spin part is composed of a single configuration of the perfect-pairing form.
The electronic energy \(E\) is computed as a function of the variables
\(\{\boldsymbol{q}_i\}\) and \(\{\rho_i\}\) at a given nuclear geometry
\(\{\boldsymbol{R}_I\}\). 
Their optimal values
\(\{\boldsymbol{q}_i^{(0)}\}\) and \(\{\rho_i^{(0)}\}\)
are determined by minimizing the energy \(E\). The 
ePES for the \(j\)-th electron
\({\cal V}_j(\boldsymbol{q})\) is constructed by fixing the variables other than
\(\boldsymbol{q}_j\) as
\begin{align}
{\cal V}_j(\boldsymbol{q})=E(\boldsymbol{q}_1^{(0)},\cdots,\boldsymbol{q}_{j-1}^{(0)},\boldsymbol{q},\boldsymbol{q}_{j+1}^{(0)},\cdots,\boldsymbol{q}_N^{(0)},
\nonumber
\\
\rho_1^{(0)},\cdots,\rho_N^{(0)};\{\boldsymbol{R}_I\}).
\label{eq:Vq}
\end{align}
The time-dependent Schr\"odinger equation for an electron
on the ePES \({\cal V}_j (\boldsymbol{q})\) 
was solved numerically.
(Note that this is not the evolution of the localized EWP
of Eq. (\ref{eq:GWP}).
The EWP are used only for the construction of ePES.)

The scheme was applied to a LiH molecule under an intense laser pulse.
The electron dynamics in LiH has been studied in many previous works
\cite{Sato2013,Sato2015,Remacle2007,Nest2008,Remacle2011,Nikodem2017,Balzer2010,Ulusoy2012}
as a prototype of simple heteronuclear molecules 
with two core and two valence electrons.
The accuracy of the VB EWP model for the ground and excited electronic
states of LiH has been examined in Ref. \cite{Ando2016}.
The static electron correlation was well described by the VB model,
and the ionic Li$^+$H$^-$ character in the ground state
was reproduced by the floating and breathing degrees of freedom
of the EWP. 
(See the insets of Figs. \ref{fig:Li2sWfn} and \ref{fig:H1sWfn}.)

To construct the potential curves
\({\cal V}_{j}(\boldsymbol{q})\),
the EWP centers
\(\boldsymbol{q}_j\) that were assigned to Li 2s and H 1s 
from their optimal position and width
were displaced along the bond direction
on the $x$-axis in the laboratory frame. 
The Li nucleus was at the origin
and the H nucleus was at $x=+2.3$ bohr \cite{Sato2015}.
For comparison, we carried out 
calculations of a free electron
under the same laser pulse.
We also computed WP dynamics with semiclassical approximation, 
where the WP widths were fixed at the optimal
values \(\{\rho_i^{(0)}\}\) and the classical equations of
motion for the WP center \(\boldsymbol{q}_j\) on
\({\cal V}_j(\boldsymbol{q})\) were solved.

In what follows, the terms \lq Li 2s electron\rq\ and \lq H 1s electron\rq\ are 
used for notional convenience,
but they differ from
the ordinary atomic orbitals clamped at the nuclear centers.
Rather, they are evolving on the ePES 
assigned to Li 2s and H 1s 
from their initial size and position.
(They may be called instead \lq weaker bound\rq\ and \lq stronger bound\rq\
valence electrons.)
The initial condition for the electronic wave function at \(t=0\) was
the form of Eq. (\ref{eq:GWP}) with the 
center \(\boldsymbol{q}_j^{(0)}\) and width
\(\rho_j^{(0)}\) optimized in the ground state without the external field:
$x^{(0)}_1=2.11$ bohr and $\rho^{(0)}_1=1.75$ bohr for the Li 2s electron and
$x^{(0)}_2=2.16$ bohr and $\rho^{(0)}_2=0.77$ bohr for the H 1s electron. 
(See the insets of Figs. \ref{fig:Li2sWfn} and \ref{fig:H1sWfn}.)

\section{Results and Discussion}
\label{sec:results}

\begin{figure}[h]
\centering
\includegraphics[width=0.46\textwidth]{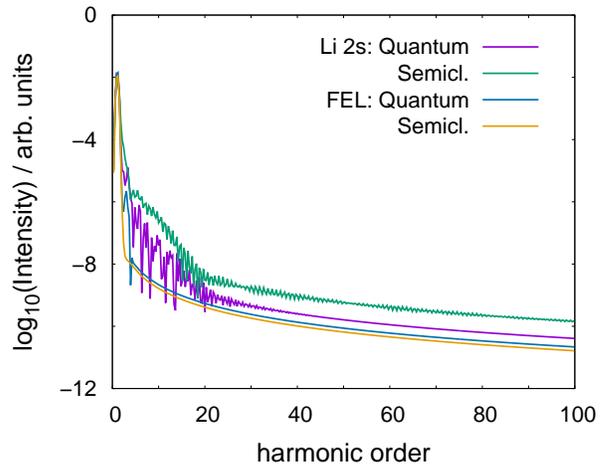}
\caption{High-harmonic generation spectra from 
an electron on the 
valence electronic potential energy curve
assigned virtually to Li 2s,
computed with full quantum and semiclassical methods,
compared with those from a free electron (FEL).}
\label{fig:Li2sHHG}
\end{figure}

\begin{figure}[h]
\centering
\includegraphics[width=0.46\textwidth]{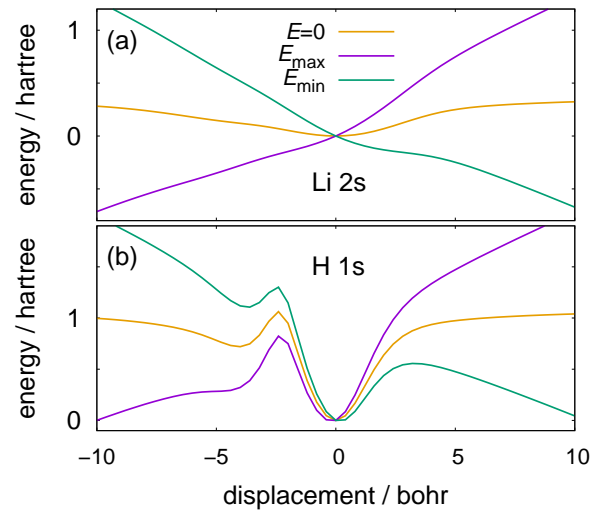}
\caption{Potential energy curves 
${\cal V}_j(x-x^{(0)}_j)$
for the valence electrons virtually assigned to
Li 2s and H 1s,
at times when the laser field intensity is zero, maximum, and minimum.}
\label{fig:vefld}
\end{figure}

We first analyze the dynamics of Li 2s electron. 
Figure \ref{fig:Li2sHHG} displays
the HHG spectrum from the quantum
and semiclassical calculations. 
The spectra from a free electron under the same laser field
are also included. All spectra
exhibit a dominant peak at the first harmonic. 
For the free electron, it is a simple consequence of
the WP motion that directly follows the laser-field
oscillation. Similarly, the dominance of the first harmonic in the Li 2s 
spectrum is comprehended to come from a free-electron-like motion due to 
the shallowness of the ePES
displayed in Fig. \ref{fig:vefld}(a),
with a
small binding energy of 0.39 hartree.

\begin{figure}[h]
\centering
\includegraphics[width=0.46\textwidth]{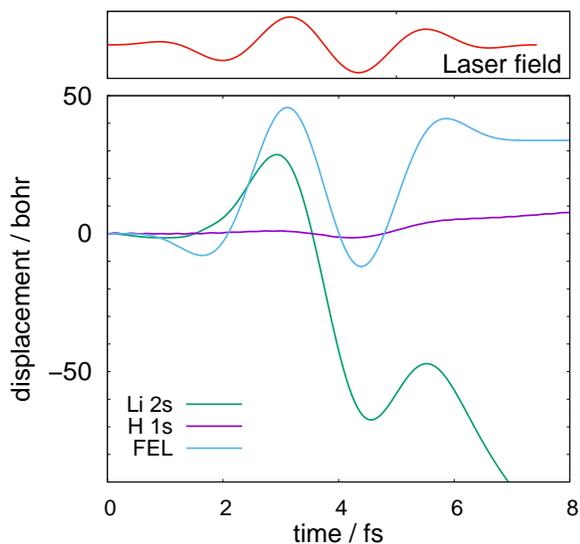}
\caption{Trajectories of the position displacement expectation value 
for electrons on the 
valence electronic potential energy curves
assigned virtually to Li 2s and H 1s
and for a free electron (FEL).
The top panel displays the time profile of the electric field intensity
of the laser pulse.}
\label{fig:LiHDplt}
\end{figure}

\begin{figure}[h]
\centering
\includegraphics[width=0.46\textwidth]{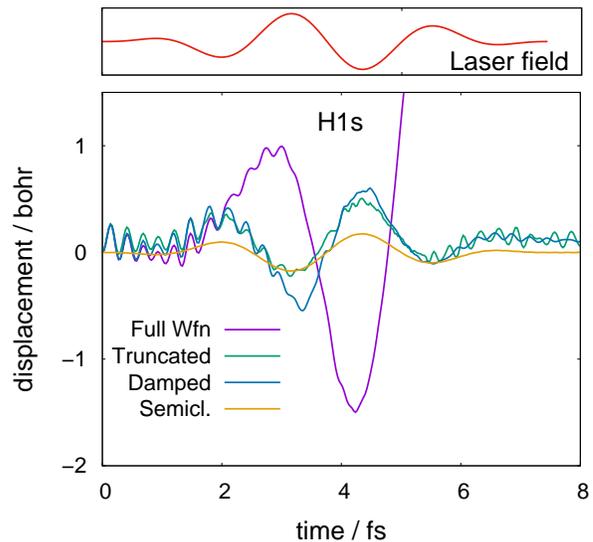}
\caption{Trajectories of the position displacement expectation value 
for an electron on the 
valence electronic potential energy curve
assigned virtually to H 1s,
computed with
full quantum, truncated, and damped wave functions 
and with semiclassical approximation.}
\label{fig:H1sDpl}
\end{figure}

\begin{figure}[h]
\centering
\includegraphics[width=0.46\textwidth]{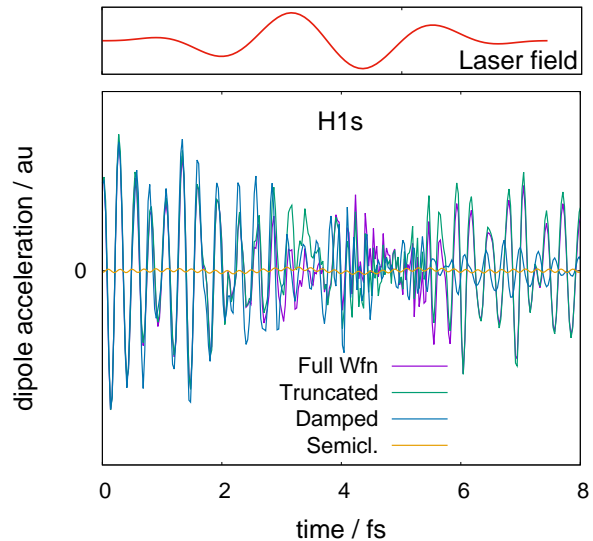}
\caption{Trajectories of the dipole acceleration 
for an electron on the 
valence electronic potential energy curve
assigned virtually to H 1s,
computed with
full quantum, truncated, and damped wave functions 
and with semiclassical approximation.}
\label{fig:H1sDac}
\end{figure}

Figure \ref{fig:LiHDplt} displays the trajectories of the position
displacement expectation value.
Similarity between the Li 2s electron and 
free electron is seen.
The amplitude of motion of the H 1s electron is much smaller,
which implies weak correlation between the Li 2s and H 1s electrons.
The trajectory of the H 1s electron will be examined later 
with Figs. \ref{fig:H1sDpl} and \ref{fig:H1sDac}.

\begin{figure}[h]
\centering
\includegraphics[width=0.46\textwidth]{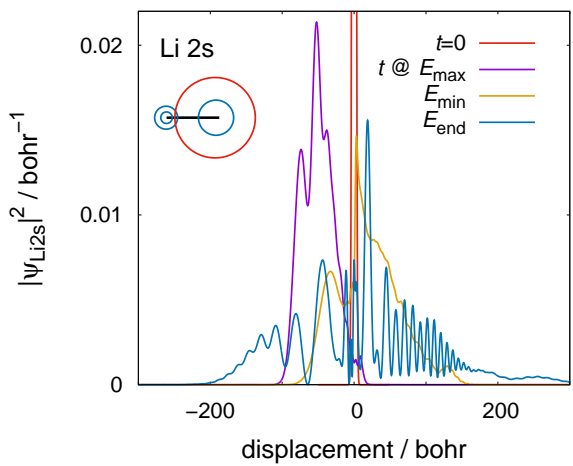}
\caption{Probability density of an electron on the 
valence electronic potential energy curve
assigned virtually to Li 2s,
at times when the laser field intensity is maximum,
minimum, and at the end of the laser pulse.
The abscissa is the displacement from the EWP center at $t=0$.
The inset at the upper-left displays circles of radius $\rho^{(0)}_i$,
the wave packet widths optimized at zero field,
with the electron wave packet assinged to Li 2s in red,
and the Li and H nuclei 
at the left and right ends of the thick black horizontal line.}
\label{fig:Li2sWfn}
\end{figure}

The probability density of the Li 2s wave function 
is plotted in Fig. \ref{fig:Li2sWfn} 
at $t=0$ and times when
the laser field intensity was the maximum (\(t = 3.2\) fs), 
minimum (\(t = 4.3\) fs), and at the end of laser pulse
(\(t = 7.5\) fs). 
(The laser pulse shape is shown in Fig. \ref{fig:LiHDplt}.)
In a few femtoseconds, the wave function broadens
to the width of 100 bohr without developing notorious peak structure. At the
end of laser pulse, the wave function acquires a complex structure.
This would be the origin of the weak peaks up to $\sim$20 harmonic orders
in Fig. \ref{fig:Li2sHHG}.

\begin{figure}[h]
\centering
\includegraphics[width=0.46\textwidth]{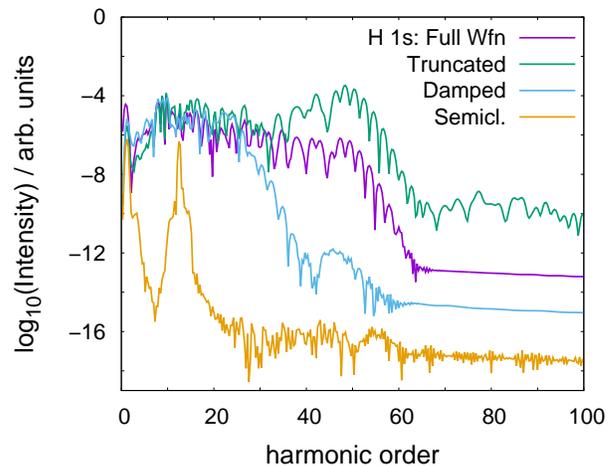}
\caption{High-harmonic generation spectra from 
an electron on the 
valence electronic potential energy curve
assigned virtually to H 1s,
computed with
full quantum, truncated, and damped wave functions 
and with semiclassical approximation.}
\label{fig:H1sHHG}
\end{figure}

\begin{figure}[h]
\centering
\includegraphics[width=0.46\textwidth]{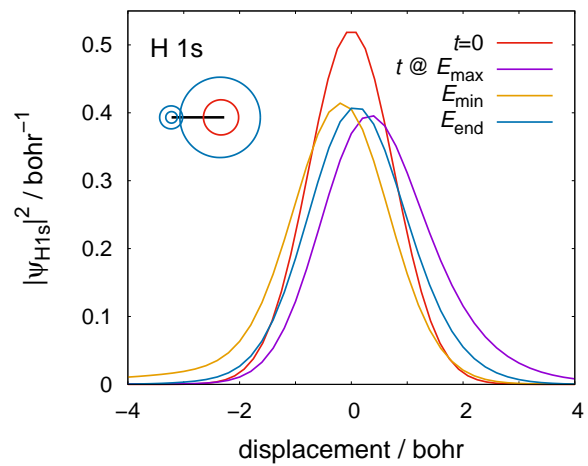}
\caption{Probability density of an electron on the 
valence electronic potential energy curve
assigned virtually to H 1s,
at times when the laser field intensity is maximum,
minimum, and at the end of the laser pulse.
The abscissa is the displacement from the EWP center at $t=0$.
The inset at the upper-left displays circles of radius $\rho^{(0)}_i$,
the wave packet widths optimized at zero field, 
with the electron wave packet assinged to H 1s in red,
and the Li and H nuclei 
at the left and right ends of the thick black horizontal line.}
\label{fig:H1sWfn}
\end{figure}

Figures \ref{fig:H1sHHG} and \ref{fig:H1sWfn} 
are the corresponding analysis for the H 1s electron.
The ePES
for H 1s electron
displayed in Fig. \ref{fig:vefld}(b),
has the
binding energy of \(\sim\)1.1 hartree, which is large enough to keep
the binding well structure under the modulation by the laser
field. Consequently, as seen in Fig. \ref{fig:H1sWfn},
the wave packet maintains the Gaussian-like shape and width. 
This seemed to imply that the semiclassical description would 
be appropriate.
However, as seen in Fig. \ref{fig:H1sHHG},
the semiclassical calculation
was unable to
reproduce the plateau and cut-off of the HHG spectrum.

\begin{figure}[h]
\centering
\includegraphics[width=0.46\textwidth]{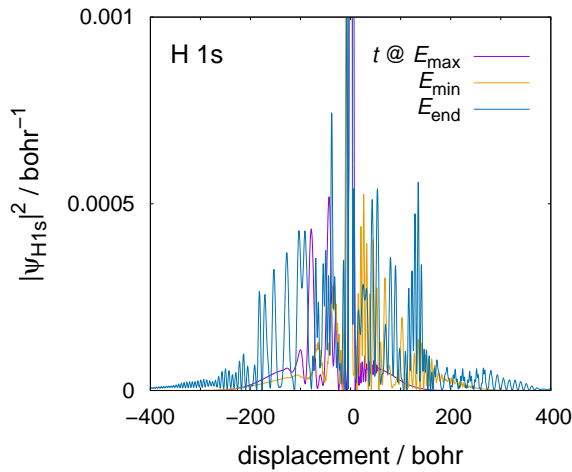}
\caption{Same as Fig. \ref{fig:H1sWfn} but in different scales of
the axes to see part of the probability densities, 
about $10^{-3}$ times smaller in amplitude and $10^{2}$
times broader in spatial range. 
}
\label{fig:H1sWfnZoom}
\end{figure}

The puzzle was resolved by looking
into fine details of the wave function. 
Figure \ref{fig:H1sWfnZoom} displays the
same probability densities 
as those in Fig. \ref{fig:H1sWfn} but in
different scales of the axes, 
about $10^{-3}$ times smaller amplitude and $10^{2}$
times broader spatial range. 
These portions of the
wave function were created via tunneling through the potential barriers
at both sides of the ePES in Fig. \ref{fig:vefld}(b).
They appear to carry the
high frequency components. 
This is confirmed by computing
the spectra with truncated and damped wave functions.
The spectrum with
the truncated wave function was computed by limiting the spatial range
of integration when computing the dipole moment by
\begin{equation}
    \mu (t) = -e \int_{x_{\rm min}}^{x_{\rm max}} x |\psi_{} (x,t)|^2 dx,
\end{equation}
that is, the wave function $\psi$ was computed normally 
but the dipole moment was computed from
the limited part of the wave function near the molecule. 
We set $x_{\rm min}=-2.4$ bohr and $x_{\rm max}=+3.2$ bohr to
include the main part of the WP in Fig. \ref{fig:H1sWfn}.
The spectrum with the damped wave
function was computed with the absorbing potential 
\cite{GonzalezLezana04} 
placed at \(x = \pm 13\) bohr
such that once a part of the wave function tunnels out of the potential well, 
it is damped out.
The difference between the truncated and damped calculations is that the
tunneled portion may return to the molecular region in the
former but not in the latter.
As seen in the spectra in Fig. \ref{fig:H1sHHG},
the damped calculation gives the plateau up to $\sim$30 harmonic orders,
but the higher region between 30 and 50 harmonic orders is suppressed.
For the truncated wave function,
the spectrum around 50 harmonic orders is rather enhanced.

The picture offered by this analysis is that the small portion of wave 
function that leaked out from the molecular binding potential 
via tunneling carries
high frequency components to yield the HHG between 30 and 50
harmonic orders. 
The HHG up to 30 harmonic orders also come from the nonlinear dynamics 
within the unharmonic potential modulated by the laser field. 

The binding energy of 1.1 hartree for the H 1s electron is 
consistent with the well-known formula for the cut-off energy 
of HHG spectra \cite{Krause1992},
\begin{equation}
    E_c = I_p + 3.17 U_p 
    \label{eq:ponderomotive}
\end{equation}
where $I_p$ is the ionization potential and $U_p$ is the ponderomotive energy.
The present parameters for the laser pulse (see Appendix) gives $U_p=0.77$ hartree.
With $I_p = 1.1$ hartree for the H 1s electron (Fig. \ref{fig:vefld}(b)), 
Eq. (\ref{eq:ponderomotive}) gives the cut-off energy of 58 harmonic orders,
consistent with the result in Fig. \ref{fig:H1sHHG}.
With the binding energy of 0.39 hartree for the Li 2s electron,
the same calculation gives 47 harmonic orders,
though the plateau and cut-off is not seen in Fig. \ref{fig:Li2sHHG}
due to the free-electron-like dynamics as discussed above.

Figure \ref{fig:H1sDpl} displays the trajectory of the position
displacement expectation value
for the H 1s electron.
In $t < 2$ fs, 
the quantum, truncated, and damped calculations agree well.
After 2 fs, when the laser field passed the minimum,
the latter two start to deviate.
The quantum result is reasonable as the electron 
is accelerated opposite
to the field direction.
Although the tunneled portion has small amplitude, it contributes
to the position expectation with large spatial 
range.
The truncated and damped results are 
from the remainder of the wave function.
The semiclassical trajectory basically follows the laser field
but with smaller amplitude.
The corresponding trajectories of the dipole acceleration
are displayed in Fig. \ref{fig:H1sDac},
in which the oscillations are seen clearer. 
The period of the prominently seen oscillation 
is $\sim$0.3 fs compared to 2.5 fs of the laser light.
The difference among the methods is small in the short time,
but becomes more apparent in the longer time,
particularly for the damped wave function.

\section{Concluding Remarks}
\label{sec:concl}

The origin of the HHG spectrum from a LiH molecule has been analyzed.
The electron dynamics 
on the weaker bound valence ePES
is the main origin of the first harmonic
but not of the higher harmonics.
In view of the spatially large amplitude oscillation of 
the electron (Fig. \ref{fig:LiHDplt}),
the three-step
model seemed to apply. 
However, the analysis indicated that the ePES the
electron is shallow such that the electron dynamics is
rather similar to that of a free electron under the oscillating laser field.
By contrast, the ePES for the other valence
electron is deep enough to hold
the major part of the wave function 
such that the oscillation amplitude under the laser field is 
as small as the molecular size,
which implies that the ionization in normal sense does not occur. 
Rather, small portion of the wave function
that leaked out of the potential well 
via tunneling
was the main origin of the HHG up to 50 harmonic orders.
Nonlinear dynamics within the potential well induced by the laser field
oscillation was another source of HHG up to 30 harmonic orders.

We note that the Li-H bond length of 2.3 bohr is shorter than the
experimental value of 3.2 bohr \cite{Orth79}.
The former was taken from Ref. \cite{Sato2015} 
which employed the soft-Coulomb potential and found
the equilibrium bond length at 2.3 bohr.
By contrast,
our model gives 3.1 bohr \cite{Ando2012}.
It would be thus intriguing to examine the effect
of bond compression and extension with our model.
In particular, the small hump of ePES 
on the negative side of the displacement 
in Fig. \ref{fig:vefld} (b)
is due to the Li nucleus, so the tunneling through the
barrier will be affected by the bond length. 
Another related issue would be inclusion of nuclear motion.
In the present framework, an approximate treatment with nuclear 
wave packets would be possible 
\cite{Kim2014,Kim2014prb,Kim2015,Kim2016}.
These are open for future examinations.

\section*{Acknowledgment}

This work was supported by KAKENHI Nos. 26248009, 26620007, and 19K22173. 

\section*{Appendix}

Here we summarize the details of 
theory and computation. 
See Ref.\cite{Ando2018} for additional information.

The electronic wave function has a form of an antisymmetrized product of
spatial and spin functions
\[
\Psi(1,\cdots,N)={\cal A}[\Phi(\boldsymbol{r}_1,\cdots,\boldsymbol{r}_N)\Theta(1,\cdots,N)]
\]
with the spatial part modeled by a product of one-electron functions
\[
    \Phi(\boldsymbol{r}_1,\cdots,\boldsymbol{r}_N)=\phi_1(\boldsymbol{r}_1)\cdots\phi_N(\boldsymbol{r}_N).
\]
The form of \(\phi_i(\boldsymbol{r})\) employed in this
work is given in Eq. (\ref{eq:GWP}). For the spin part,
we employ a single
configuration of the perfect-pairing form
\[
\Theta = \theta(1,2)\theta(3,4)\cdots\theta(N-1,N)
\]
in which \(\theta(i,j)=(\alpha(i)\beta(j)-\beta(i)\alpha(j))/\sqrt{2}\).

The time-dependent electric field of the laser pulse has a form
\[
{\cal E}(t)={\cal E}_0\sin(\omega_0t)\sin^2(\pi t/\tau)
\quad
(0 \le t \le \tau)
\]
in the direction parallel to the Li-H bond. 
The frequency \(\omega_0\) corresponds
to the wavelength 750 nm. The duration \(\tau\) is for three optical
cycles, \(\tau=3(2\pi/\omega_0)\simeq\) 7.51 fs. The field intensity is
\({\cal E}_0=5.5\times 10^8\) V/cm which corresponds to the laser
intensity of \(4.0\times 10^{14}\) W/cm\(^2\).
These parameter values were taken from Ref. \cite{Sato2015}. 
The positive value of ${\cal E}$ corresponds to the electric field
in the direction from Li to H nuclei.

The length of the simulation box was taken to be 1200 bohr, with the
transmission-free absorbing potential
\cite{GonzalezLezana04} of 120 bohr length at both ends.
The wave functions were propagated with the Cayley's hybrid scheme
\cite{Watanabe00} with the spatial grid length of 0.2
bohr and the time-step of 0.01 au (\(\sim\)0.24 as). The norm of the
wave function stayed unity with the deviation less than 10\(^{-7}\)
throughout the simulation. 
The semiclassical calculation was by the frozen Gaussian method
with the fixed width \(\rho_j^{(0)}\).
The derivatives of the potentials \({\cal V}_j(\boldsymbol{q})\) were
computed by the spline interpolation.
The HHG spectra were computed from the
Fourier transform of the dipole acceleration dynamics.

\end{document}